\documentstyle[preprint,aps,floats,epsfig]{revtex}

\newcommand{\be}{\begin{equation}}
\newcommand{\ee}{\end{equation}}
\newcommand{\bea}{\begin{eqnarray}}
\newcommand{\ea}{\end{eqnarray}}
\newcommand{\bml}{\begin{mathletters}}
\newcommand{\eml}{\end{mathletters}}

\begin{document}

\tighten

\preprint{DCPT-02/35}
\draft




\title{Critical phenomena of gravitating monopoles in the spacetime
of a global monopole}
\renewcommand{\thefootnote}{\fnsymbol{footnote}}
\author{ Yves Brihaye\footnote{Yves.Brihaye@umh.ac.be}}
\address{Facult\'e des Sciences, Universit\'e de Mons-Hainaut,
 B-7000 Mons, Belgium}
\author{Betti Hartmann\footnote{Betti.Hartmann@durham.ac.uk}}
\address{Department of Mathematical Sciences, University
of Durham, Durham DH1 3LE, U.K.}
\date{\today}
\setlength{\footnotesep}{0.5\footnotesep}

\maketitle
\begin{abstract}
We present a  numerical study of critical phenomena (including the
Lue-Weinberg phenomenon) arising for gravitating monopoles
in a global monopole spacetime. The equations 
of this model have been recently studied by Spinelly et al.
in a domain of parameter space away from the critical
points. 
\end{abstract}


\renewcommand{\thefootnote}{\arabic{footnote}}

\section{Introduction}
Topological defects \cite{vilenkin} arise in spontaneously broken field theories 
with non-trivial topological structure
of the vacuum. If the symmetry broken is global,
the defects have infinite energy and thus do not
posses a particle-like behaviour.
One way to get rid of this problem is to  introduce gauge symmetries,
the most popular example being the 't Hooft-Polyakov magnetic monopole 
\cite{thooft}.
Although initially constructed within an $SU(2)$ gauge field theory, 
monopoles are predicted by all Grand Unified Theories (GUTs)
which contain an unbroken $U(1)$ symmetry. Since we observe this unbroken
$U(1)$ symmetry in the universe, this leads to the so-called "monopole-problem",
the most popular solution to which is the scenario of inflation \cite{guth}.

The coupling of field theories to gravity  enriches their classical spectrum. 
For example, the Lagrangian describing a triplet of self-interacting
scalar fields invariant under a global $O(3)$ transformation
has no finite energy solutions. However, the incorporation of 
gravity leads to  global monopoles 
\cite{vile,harari} which have particle-like structure.

The most striking feature of classical field theory solutions
coupled to gravity is the pattern of bifurcations found e.g. for the
gravitating monopoles merging into black hole solutions \cite{weinberg,bfm}.

Recently, a model involving both a $SO(3)$ triplet of Higgs 
fields as well as an $O(3)$ triplet of Goldstone field was considered in \cite{spine}. 
Coupling the Lagrangian of this model to gravity, the authors 
were able to construct finite mass solutions incorporating
both the gravitating monopole of \cite{bfm} and the global monopole of
\cite{vile}.
Because many features of these solutions were left open, we reconsider here the
classical equations and put the emphasis on several unsolved questions, namely~:
(i) how does the topological defect emerge from the purely gravitating magnetic
monopole, (ii) what is the domain of existence of the solutions in the space
of the parameters, (iii) do the solutions bifurcate into black holes solutions,
(iv) do these features persist for large values of the self-interacting 
coupling constant. These questions are worth studying because it is known that
the global monopole has a much stronger gravitational field at large distances
as compared to that of the local monopole. The reason for this is that the space-time of
the global monopole is not asymtotically flat, while that of the gauged monopole is.
However, at short distances the gravitational effects of the global monopole
are weak as long as its mass is much smaller than the Planck mass \cite{vilenkin}.
On the other hand, it is well known that
local monopoles stop to exist when their radius becomes comparable to the corresponding
Schwarzschild radius, which implies that  the mass of the monopoles is of order of the
Planck mass \cite{weinberg,bfm}. Since we are studying critical phenomena of these
monpoles, the mass of both the local and the global monopole are of the order
of magnitude of the Planck mass and the argument that the short distance gravitational field
of the global monopole is weak doesn't hold anylonger. One might thus expect 
significant changes of the critical behaviour of the local monopole in the spacetime of a global
one.

In order to answer these questions, we consider the model of \cite{spine}
for generic values of the expectation values of the local 
Higgs fields and of the global Goldstone fields. The Lagrangian, 
the spherically symmetric Ansatz and the classical
equations are presented in Sect. II. The numerical results are 
discussed in Sect. III
and illustrated by means of five figures. We give our conclusions in Sec. IV.

\section{The Model}
\subsection{The Lagrangian}
We consider the following  Lagrangian density \cite{spine}~:
\be
\label{lmat}
{\cal L}_M = -\frac{1}{4}F_{\mu \nu} ^a F^{\mu \nu,a} 
- \frac{1}{2}D_{\mu} \phi^a D^{\mu} \phi^a
- \frac{1}{2}\partial_{\mu} \xi^a \partial^{\mu} \xi^a  - V(\phi^a, \xi^a) 
\end{equation}
with the potential
\be
    V(\phi^a, \xi^a) = \frac{\lambda}{4}(\phi^a\phi^a- \upsilon^2)^2
           + \frac{ \lambda_g}{4}(\xi^a\xi^a- \eta^2)^2   \ , \ 
\end{equation}
the covariant derivative of the  Higgs triplet $\phi^a$, $a=1,2,3$ 
\begin{equation}
D_{\mu} \phi^a = \partial_{\mu}\phi^a - e \varepsilon_{abc} A_{\mu}^b \phi^c
\end{equation}
and the field strength tensor
\begin{equation}
F_{\mu \nu} ^a = \partial_{\mu} A_{\nu}^a - \partial_{\nu} A_{\mu}^a
                - e \varepsilon_{abc} A_{\mu}^b A_{\nu}^c \ . \
\end{equation}
This model contains two triplets of scalar fields: the Higgs fields
$\phi^a$ with vacuum expectation value $\upsilon$
and self-coupling $\lambda$ and the Goldstone fields $\xi^a$, ($a=1,2,3$)
with vacuum expectation value $\eta$ and self-coupling $\lambda_g$.
It is invariant under a $SO(3)_{local} \times O(3)_{global}$
transformation~:
\begin{equation}
\label{symmetry}
(\phi^a)' (x) = R_{ab}(x) \phi^a(x) \quad , \quad 
(\xi^a)'(x) = {\tilde R}_{ab} \xi^a(x)
\end{equation}
where the "$R$-part" of the symmetry is gauged by means of the 
$SO(3)$ Yang-Mills field $A^a_{\mu}$ with gauge coupling $e$,
while  the "$\tilde R$-part"
denotes a global transformation.
Note that for the Lagrangian (\ref{lmat}) the two scalar
triplets are
decoupled. The interaction between these fields 
will be carried out through the coupling to gravity.
We thus consider the following action~:
\begin{equation}
\label{action}
           S =\int \left( \frac{R}{16\pi G}+  {\cal L}_M \right) \sqrt{-g} d^4 x
\end{equation}
where $G$ is Newton's constant, $g$ denotes the determinant of the
metric tensor $g_{\mu\nu}$ and $R=g^{\mu\nu}R_{\mu\nu}$ is the Ricci scalar.
In this paper, we are carrying
out a detailed numerical study of the classical, 
spherically symmetric solutions 
of the system of equations which arise from the variation of (\ref{action}).

\subsection{Spherically symmetric Ansatz}

For the metric, the spherically symmetric Ansatz
in Schwarzschild-like coordinates reads \cite{weinberg,bfm,bhk}~:
\begin{equation}
ds^{2}=g_{\mu\nu}dx^{\mu}dx^{\nu}=
-A^{2}(r)N(r)dt^2+N^{-1}(r)dr^2+r^2 (d\theta^2+\sin^2\theta
d\varphi^2)
\label{metric}
\ , \end{equation}
For the gauge, Higgs and Goldstone fields, we use 
the purely magnetic hedgehog Ansatz
\cite{thooft,vile,spine}~:
\begin{equation}
{A_r}^a={A_t}^a=0
\ , \end{equation}
\begin{equation}
{A_{\theta}}^a= \frac{1-u(r)}{e} {e_{\varphi}}^a
\ , \ \ \ \ 
{A_{\varphi}}^a=- \frac{1-u(r)}{e}\sin\theta {e_{\theta}}^a
\ , \end{equation}
\begin{equation}
{\phi}^a= \upsilon  h(r) {e_r}^a    \quad , \quad 
 {\xi}^a = \upsilon  f(r) {e_r}^a 
\ . 
\end{equation}
We introduce the following dimensionless variable and coupling constants~:
\begin{equation}
     x = e \upsilon r \quad , \quad 
     \alpha^2 = 4 \pi G \upsilon^2 \quad , \quad 
     \beta^2=\frac{\lambda}{e^2} \quad , \quad 
     \beta_g^2=\frac{\lambda_g}{e^2} \quad , \quad 
     q = \frac{\eta}{\upsilon} 
\end{equation}
The ratio between the radius of 
the local monopole core $r_l\propto (ev)^{-1}$ and
the global monopole core
$r_g\propto(\sqrt{\lambda_g}\eta)^{-1}$ can be given in terms of
these quantities~:
\begin{equation}
\frac{r_l}{r_g}\propto\frac{\sqrt{\lambda_g}\eta}{e \upsilon}=\beta_g q
\end{equation}
Note that the notation used here corresponds to the one used in \cite{bhk}
and differs from the one in \cite{spine}. To avoid confusion,
we stress here the crucial differences~:
\begin{center}
\cite{spine}:     $\alpha^2 = 1 - 8 \pi G \eta^2$ \ ,  \ \  $g_{rr}=A$  
\ , \  \ $g_{tt}=B$ \\
\cite{bhk}:    $\alpha^2 = 4 \pi G \upsilon^2$ \ , \ \ 
   $g_{rr}=N^{-1}$ \ ,  \  \ $
g_{tt}=A^2 N$
\end{center}
\subsection{Classical field equations}
Varying (\ref{action}) with respect to the metric fields gives 
the Einstein equations which can be combined to give two first order
differential equations for $A$ and $\mu$:~
\begin{equation}
\label{a}
A' = \alpha^2 A x 
(\frac{2}{x^2}(u')^2  + (h')^2 + (f')^2) 
\end{equation}
\begin{equation}
\label{mu}
\mu' = \alpha^2 x^2 (({\cal U} - \frac{q^2}{x^2}) + N {\cal K}) 
\end{equation}
with the abbreviations
\begin{eqnarray}
&{\cal K} &= \frac{1}{2}(\frac{2}{x^2}(u')^2  +(h')^2 +(f')^2) \ , \  \\ 
&{\cal U} &= \frac{(u^2-1)^2}{2 x^4} + h^2 \frac{u^2}{x^2} + \frac{f^2}{x^2}
+ \frac{\beta^2}{4}(h^2-1)^2 + \frac{\beta_g^2}{4}(f^2-q^2)^2  \ , \
\end{eqnarray}
and $N$ and $\mu$ are related as follows:~
\begin{equation} 
\label{nmu}
N(x) = 1 - 2\alpha^2 q^2 - 2 \frac{\mu(x)}{x} \ . \
\end{equation}
Variation with respect to the matter fields yields the Euler-Lagrange
equations, which for our model are a set of three second order differential equations:~
\begin{eqnarray}
&(A N u')' &= A( \frac{u(u^2-1)}{x^2} + h^2 u) \ ,  \\
\\
&(x^2 A N h')' &= A( 2 h u^2 + \beta^2 x^2 h(h^2-1)) \ , \\
\\
&(x^2 A N f')' &= A( 2 f +  \beta_g^2 x^2 f(f^2-q^2))  \ . \
\label{feq}
\end{eqnarray}
The prime denotes the derivative with respect to $x$.\
In order to solve the system of equations uniquely,
we have to introduce $8$ boundary conditions, which we choose to be~: 
\begin{equation}
     \mu(0) = 0 \quad , \quad u(0) = 1 \quad , \quad h(0) = 0 \quad , \quad f(0) = 0
\end{equation}
\begin{equation}
     A(\infty) = 0 \quad , \quad u(\infty) = 0 \quad , 
     \quad h(\infty) = 1 \quad , \quad f(\infty) = q
\end{equation}
Note that close to the origin, $\mu(x\rightarrow 0)
\approx -\alpha^2 q^2 x$ such that
$N(x\rightarrow 0) \rightarrow 1$. Note further that while the decay of the Higgs field function $h$
for $x \rightarrow \infty$  depends on $\beta$, 
the decay of the Goldstone field   function $f$ doesn't depend on $\beta_g$~: 
\begin{equation}
(h-1)\approx \exp(-\sqrt 2 \beta x) \quad , \quad (f-q)\approx c/x^2 \ \ \ \ \ \ 
{\rm for} \ \ x\rightarrow\infty \ . \
\end{equation}

The dimensionless mass of the solution is determined by the asymptotic 
value $\mu(\infty)=\mu_{\infty}$  of the function $\mu(x)$ and is given
by $\mu_{\infty}/\alpha^2$.

\section{Numerical results}
\subsection{Gravitating monopoles: $q=0$}

For $q=0$, no global symmetry breaking takes 
place and equation (\ref{feq}) is trivially solved by $f=0$.
In this limit, the remaining equations are those of
the gravitating magnetic 
monopole studied in \cite{bfm}. 
For completness, we briefly recall the main properties of these solutions.
The  $\alpha=0$ limit corresponds to flat space with $N(x)=A(x)=1$ and
thus the 't Hooft- Polyakov \cite{thooft}
monopole is recovered. For non-vanishing  $\alpha^2$, space-time is curved
and the gravitating monopole arises smoothly from its flat space limit.
While the mass of the gravitating monopole given (in our rescaled coordinates
) by $\mu_{\infty}/\alpha^2$ decreases as expected, the ratio 
$\mu_{\infty} /\alpha$ increases gradually from zero to one.
In particular, the function $N(x)$ develops a minimum $N_m$ at $x = x_m$
which becomes deeper for increasing $\alpha$.
At a $\beta$-dependent critical value of $\alpha$ 
(e.g. $\alpha_{cr} (\beta=0) \approx 1.385$,
$\alpha_{cr} (\beta=1) \approx 1.145$)
a degenerate horizon forms with $N_m \rightarrow 0$ and
the non-abelian solution ceases to exist.
It bifurcates into an embedded extremal Reissner-Nordstr\"om (RN)
black hole solution with magnetic charge $P=1$ and horizon $x_h$~:
\be
            N_{RN}(x) = 1 - \frac{2\mu_{\infty}}{x} +
 \frac{\alpha^2}{x^2} \quad , \quad A_{RN}(x)=1 \quad , \quad \mu_{\infty}=x_h=\alpha
\ee
For $0 \leq \beta \leq 0.757$, the solutions exist up to a maximal value
$\alpha_{max}$ with $N_m(\alpha_{max}) \neq 0$ and reach their critical
solution on a second branch of solutions with $\alpha_{cr} < \alpha_{max}$. 
This second branch disappears for $\beta > 0.757$ and $\alpha_{cr}=\alpha_{max}$.
For intermediate values of $\beta$, i.e. for
$\beta \ge \beta_{tr}=7.15$, however, the pattern of reaching the 
critical solution changes \cite{lue,bhk2}. Lue and Weinberg \cite{lue}
observed that for large enough $\beta$ and 
increasing $\alpha$, a second local minimum of $N(x)$ starts to form
which is located between the origin and the minimum which corresponds to
the Reissner-Nordstr\"om horizon. Eventually, the inner minimum dips down much
quicker than the outer one, such that the critical solution is an extremal,
non-abelian black hole with mass less than that of
the corresponding extremal RN solution. 
For $\beta$ close to $\beta_{tr}$, the outer
minimum is already quite pronounced at the moment the inner minimum starts
to form \cite{bhk2}.

\subsection{Gravitating monopoles: $q\neq 0$}
One of the main goals of this paper is to study how the critical phenomena
described in the previous section   
change in the presence of a global monopole, i.e.
for $q > 0$. \

This is illustrated in FIGs. 1 and 2 for  
$\beta = \beta_g = 1$ and $\alpha = 0.6$. FIG. 1
demonstrates the evolution of the mass function $\mu(x)$ 
for different values of the parameter $q$.
$\mu(x)$ develops a  local, negative valued minimum and
its asymptotic value decreases with $q$,
in particular $\mu_{\infty}$ becomes negative for $q >  0.7$.
The evolution of the function $N(x)$ is presented in
FIG. 2. The difference $N(\infty) - N_m$, where $N_m$ denotes
the local minimum of $N(x)$, decreases for increasing
$q$ and the function becomes monotonically decreasing for $q > 0.8$.
The case of equal vacuum expectation values $q=1.0$ was studied in 
\cite{spine}. Note that the solutions cease to exist for
$q > 1/ (\sqrt 2 \alpha) \approx 1.178$.\

The natural question to raise now is in which domain
of the $\alpha$, $\beta$, $\beta_g$, $q$ hyperspace monopole solutions
exist. For the moment,
we limit our analysis to the domain of existence in the  
$\alpha-q-$plane for fixed $\beta$, $\beta_g$.   

First, we analyse the evolution of the 
solutions for fixed small values of  $q\leq q_{tr}(\beta,\beta_g)$ and 
increasing $\alpha$.
We find a very similar picture as
in the  $q=0$ case~:
the function $N(x)$ develops a minimum which approaches zero for 
$\alpha \approx \alpha_{cr}(q,\beta,\beta_g)$. Our numerical analysis suggests
that the critical value of $\alpha$ increases for increasing $q$ \cite{numerics}, e.g. for
$\beta=\beta_g=1.0$ we find~:
\be
\alpha_{cr}(0.0,1.0,1.0) \approx 1.145 \quad , \quad
\alpha_{cr}(0.2,1.0,1.0) \approx 1.145  \quad , \quad
\alpha_{cr}(0.6,1.0,1.0) \approx 1.175  
\ee
We further find that the critical value of $\alpha$ depends only little
on $\beta_g$, e.g. for $q=0.1$ and $\beta=1.0$ we obtain~:
\be
\alpha_{cr}(0.1,1.0,1.0) \approx 1.145 \quad , \quad
\alpha_{cr}(0.1,1.0,5.0) \approx 1.146  \quad , \quad
\alpha_{cr}(0.1,1.0,10.0) \approx 1.147  
\ee
For small values of $\beta$, we find in analogy to \cite{bfm} 
that the solutions exist up to a maximal value of 
$\alpha$, $\alpha_{max}(q,\beta,\beta_g)$ with $N_m \neq 0$ and that from there
a second branch of solutions exists up to $\alpha_{cr} <  \alpha_{max}$
with $N_m$ reaching zero at $\alpha=\alpha_{cr}$. Both, $\alpha_{max}$
and $\alpha_{cr}$ increase with $q$, e.g. for $\beta=0.0$, $\beta_g=1.0$,
the values are~:
\be
\alpha_{cr}(0.0,0.0,1.0) \approx 1.385 \quad , \quad
\alpha_{cr}(0.3,0.0,1.0) \approx 1.399  \quad , \quad
\alpha_{cr}(0.4,0.0,1.0) \approx 1.434  
\ee
For $q > q_{tr}(\beta,\beta_g)$
the scenario is quite different as can be guessed from FIG.~2. 
Indeed, when the
expectation value of the global Goldstone field becomes large,
the function $N(x)$ decreases monotonically from $N(0) = 1$
to its asymptotic value $N(\infty) = 1 - 2 \alpha^2 q^2$. 
No local minimum develops  and the solution just stops existing
because the asymptotic value  $N(\infty)=1-2 q^2 \alpha^2$ of $N(x)$
itself becomes negative. 
The domain of existence
of solutions in the $\alpha-q-$ plane is presented 
in FIG.~3. This FIG. suggests
clearly that $q_{tr}(\beta,\beta_g)$ decreases with decreasing $\beta$, which is
indeed what our numerical simulations confirm.

In order to understand the pattern of reaching the critical solution, we show
in FIG. 4. 
the critical solution for $q=0.2$. Our numerical
results suggest that a degenerate horizon forms
at  $x_h \approx 1.148$. For $0 > x > x_h$, the Goldstone 
field function $f(x)$ vanishes, while all other functions are non-trivial.
For $x > x_h$ in contrast, $u(x)\equiv 0$, $h(x)\equiv 1$, while
the Goldstone field function $f(x)$ and the metric functions $N(x)$ and 
$A(x)$ remain non-trivial
in this region. We can make a rough approximation to explain this result
analytically as follows~: 
at and just outside the horizon, we can treat the Goldstone field 
as roughly vanishing $f(x)\approx 0$. Thus equation (\ref{mu}) becomes
\begin{equation}
\mu'=\alpha^2 x^2 (\frac{1}{2 x^4} - \frac{q^2}{x^2} + \beta_g q^4) \  \
{\rm for} \ \ x\approx x_h
\end{equation}
For $q=0$ the solution is clearly the RN solution. For $q\neq 0$, 
however, we find  using (\ref{nmu}):
\begin{equation}
N(x)=1+\frac{\alpha^2}{x^2}+\frac{C}{x}-\frac{2}{3} \alpha^2 q^4 \beta_g^2 x^2
\end{equation}
where $C$ is an integration constant. This solution has degenerate
horizons with $N(x_h)=N'(x_h)=0$ at
\begin{equation}
\label{horizon}
x^{(1,2,3,4)}_h=
\pm \frac{1}{2\alpha q^2 \beta_g}\sqrt{1\pm \sqrt{1-8\alpha^4 q^4 \beta_g^2}  }
\end{equation}
For our choice of parameters in FIG. 4, one of these four horizons is
$x_h\approx 1.148$. This is exactly equal to the horizon we find in our numerical
calculations. This approximation is, of course, only valid close to the horizon. 
Away from the horizon, the solution is non-abelian. We thus observe a "black hole
inside a global monopole". Outside the core of the global monopole, where $f$ has reached
its vacuum expectation value $f(x)=q$, (\ref{mu}) reduces to the
equation for the RN case and using (\ref{nmu}) we obtain for $N(x)$ the metric function
of an extremal RN black hole with deficit angle $1- 2 \alpha^2 q^2$ :
\begin{equation}
N_{RN}(x) = 1-2 \alpha^2 q^2 - \frac{2\mu_{\infty}}{x} + \frac{\alpha^2}{x^2} \ . \
\end{equation}
The mass of this solution is given by
$\mu_{\infty}/\alpha^2=\alpha^{-1}\sqrt{1-2\alpha^2
q^2}$. For our choice of parameters in FIG. 4, our numerical results
indicate that indeed at the critical value of $\alpha=1.145$, this mass is obtained.

Finally, we demonstrate in FIG.~5 that the so-called
Lue-Weinberg (LW) phenomenon observed previously only in the
gravitating monopole case for large enough values of $\beta$ \cite{lue,bhk2} persists
in the presence of a global monopole. We show the metric function $N(x)$ for
$\beta=15$, $\beta_g=1.0$ and $q=0.5$. Apparently, with increasing $\alpha$, first 
a RN type horizon forms at roughly (using (\ref{horizon})) $x\approx 0.82$. At 
$\alpha=0.7975$, a second minimum starts to develop, which dips down much quicker than
the RN type minimum. At $\alpha_{cr}\approx 0.79895$, this inner minimum has reached
zero, while the outer one is still greater than zero. Thus, an extremal
non-abelian black hole has formed. This scenario is similar to the one
observed in the model without a global Goldstone field. Comparing the
critical value of $\alpha$, we find that $\alpha_{cr}(q,\beta=15.)$ decreases slightly
with increasing $q$. We find $\alpha_{cr}(0.0)=0.80017$ 
and $\alpha_{cr}(0.5)=0.79895$.

\section{Conclusions}
We have studied gravitating magnetic monopoles in the spacetime of
a global monopole and have put emphasis on the study of
critical phenomena. We find that the solutions merge into
extremal black hole solutions representing
"black holes inside a global monopole" for a choice of parameters
where the radius of the core of the local monopole is smaller than
that of the global one (i.e. for small values of the product $\beta_g q$).
The critical value of $\alpha$ is increasing with increasing $q$.
For intermediate values of the Higgs boson mass, we find that the
Lue-Weinberg phenomenon persists in the presence of a global monopole
with the critical value of $\alpha$ decreasing with increasing 
$q$ for this phenomenon.

It is remarkable that many properties of the gravitating monopole persist in the
presence of a global monopole. Of course, the model studied here involves
many parameters and we limited our analysis to some particular cases. We believe though
that all qualitative properties of the solutions are exhibited in our results.

It would be interesting to study the corresponding black hole solutions of this model,
especially to investigate how the domain of existence in the $\alpha$-$x_h$-plane
chances in the presence of a global monopole.

Axially symmetric $SU(2)$ monopoles in curved space have been studied in \cite{hkk}.
It was found that in contrast to flat space, an attractive phase can exist for
specific choices of the coupling constants. It is left as future work to study the
influence of the global monopole on the attraction between like-charged gauged
monopoles.\\
\\

{\bf Acknowledgements}
B. H. was supported by the EPSRC.

\newpage


\begin{figure}\centering\epsfysize=20cm
\mbox{\epsffile{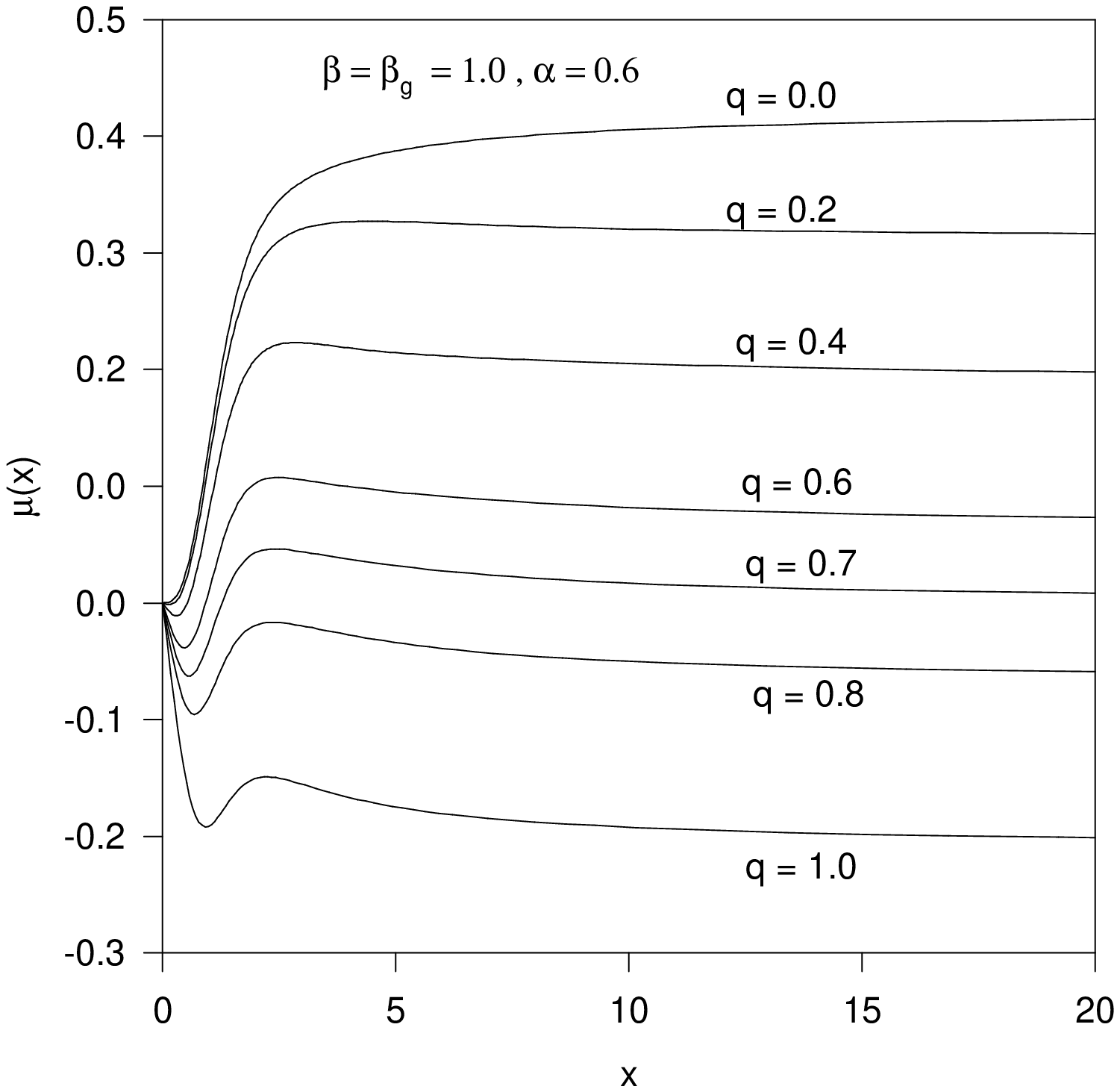}}
\caption{
The mass function $\mu(x)$ is presented for $\beta=\beta_g=1.0$,
$\alpha = 0.6$ and for several values of $q$ as a function of the dimensionless
coordinate $x=evr$.}
\end{figure}

\begin{figure}\centering\epsfysize=20cm
\mbox{\epsffile{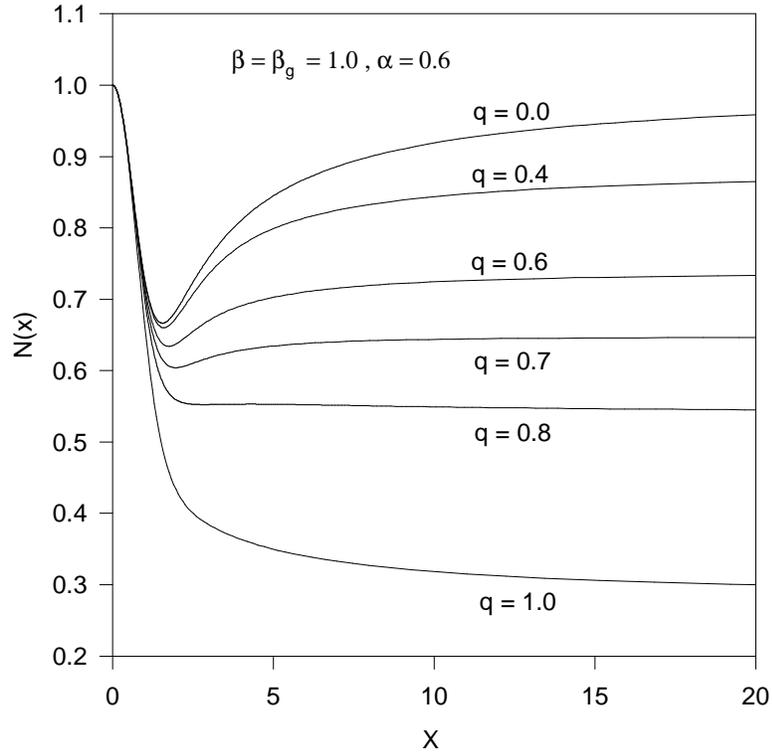}}
\caption{
Same as FIG.~1 for the metric function
$N(x)$.}
\end{figure}
\newpage
\begin{figure}\centering
\epsfysize=20cm
\mbox{\epsffile{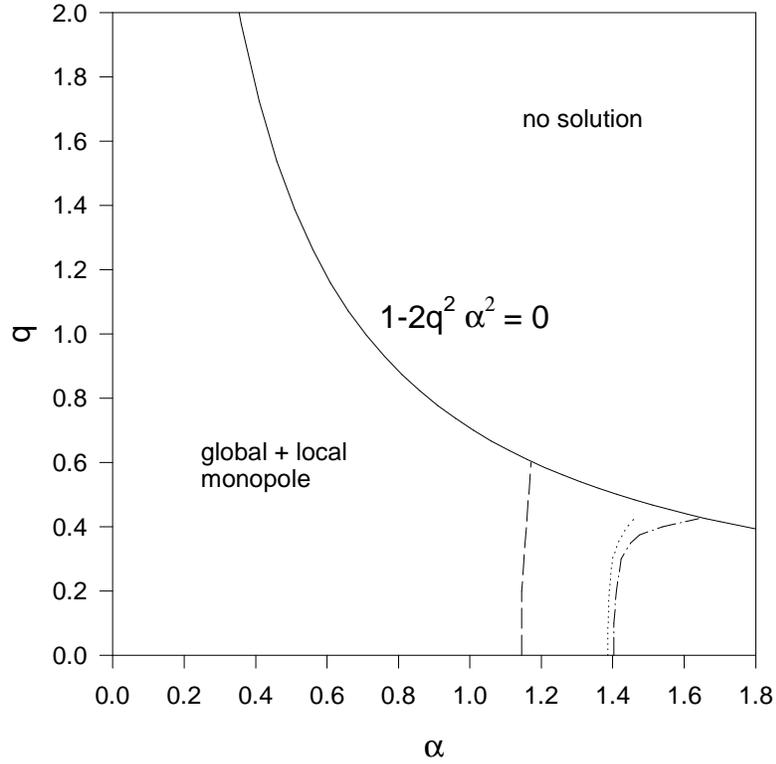}}
\caption{
The  domain of existence of the solutions in the $\alpha-q-$ plane is shown
for $\beta_g =\beta= 1.0$ (dashed)  and  $\beta_g=1.0$, $\beta=0$ 
($\alpha_{max}$~: dotted-dashed, $\alpha_{cr}$~: dotted). The solid
line represents $1-2q^2\alpha^2=0$. The solutions exist below the solid line
and to the left of the corresponding dashed, dotted-dashed, dotted line, respectively.}
\end{figure}
\newpage
\begin{figure}\centering\epsfysize=20cm
\mbox{\epsffile{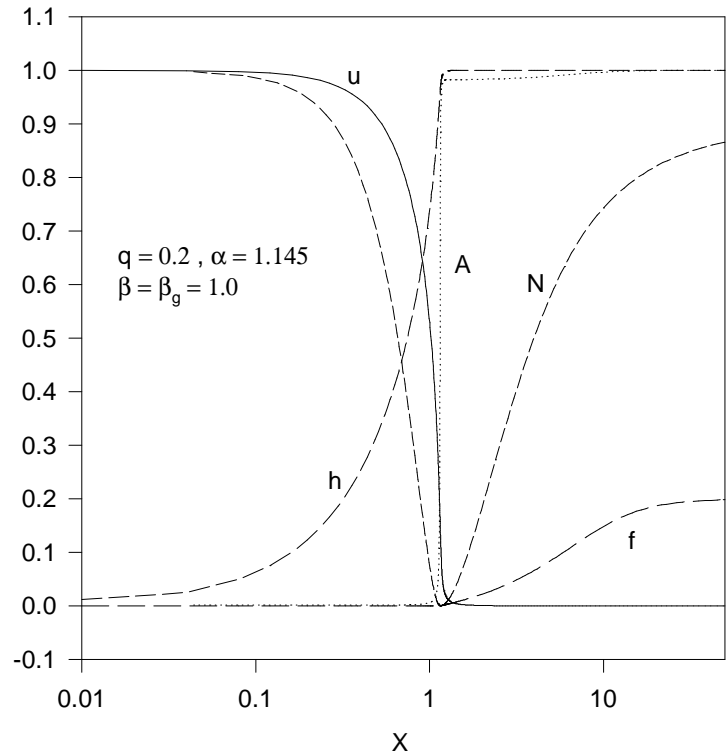}}
\caption{
The  functions $N, A, u, h, f$ are presented for $\beta=\beta_g=1.0$,
$q=0.2$ and
$\alpha \approx \alpha_{cr} \approx 1.145$  as functions
of the dimensionless coordinate $x=evr$.}
\end{figure}
\newpage

\begin{figure}\centering
\epsfysize=20cm
\mbox{\epsffile{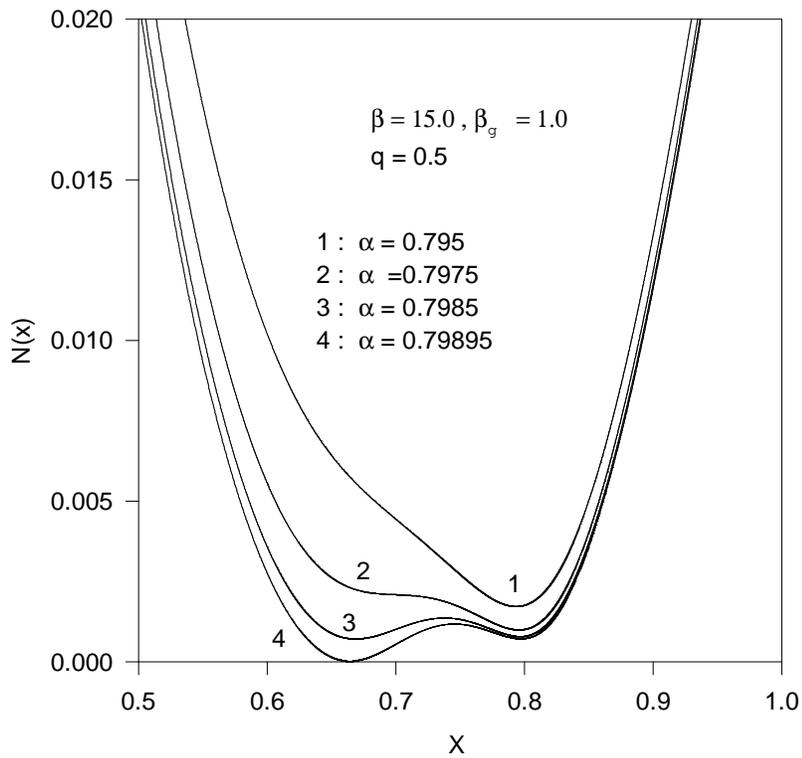}}
\caption{
The metric function $N(x)$ is shown for
$\beta=15$, $\beta_g=1.0$, $q=0.5$ and four values of $\alpha$, especially
$\alpha\approx \alpha_{cr}=0.79895$ as function of the
dimensionless coordinate $x=evr$.}

\end{figure}

\begin{thebibliography}{99}
\bibitem{numerics}
It should be stressed that the numerical integration of the equations becomes
more and more involved for increasing $q$.
\end{thebibliography}
\end{document}